\documentclass[%
reprint,
 amsmath,amssymb,
 aps,
 prb,
]{revtex4-2}

\usepackage{graphicx}% Include figure files
\usepackage{dcolumn}% Align table columns on decimal point
\usepackage{bm}% bold math
\usepackage{hyperref}% add hypertext capabilities
\usepackage{comment}
\usepackage{tikz}
\hypersetup{colorlinks=true, linkcolor=blue, filecolor=blue, urlcolor=black, citecolor=blue}
\begin{document}

\preprint{APS/123-QED}

\title{Cubic magnetic anisotropy in $B$20 magnets: Interplay of anisotropy and magnetic order in Fe$_{1-x}$Co$_{x}$Si}

\author{G. G\"odecke$^1$}
\email{g.goedecke@tu-braunschweig.de}
\author{A. O. Leonov$^{2,3}$}
\email{leonov@hiroshima-u.ac.jp}
\author{J. Grefe$^1$}
\author{S. S\"ullow$^1$}
\author{D. Menzel$^1$}

\affiliation{$^1$Institut f\"ur Physik der Kondensierten Materie, TU Braunschweig, D-38106 Braunschweig, Germany
}
\affiliation{$^2$Department of Chemistry, Faculty of Science, Hiroshima University Kagamiyama, Higashi Hiroshima, Hiroshima 739-8526, Japan
}
\affiliation{$^3$International Institute for Sustainability with Knotted Chiral Meta Matter (WPI-SKCM2), Hiroshima University, 1-3-1 Kagamiyama, Higashi-Hiroshima, Hiroshima 739-8526, Japan}

\date{\today}% It is always \today, today,

\begin{abstract}
The metallic systems MnSi and Fe$_{1-x}$Co$_{x}$Si are known to feature a generic magnetic phase diagram primarily determined by the isotropic exchange and Dzyaloshinskii-Moriya interactions. However, additional weaker anisotropies, lowest in the hierarchy of energy scales, play a crucial role: they determine the relative order of phases in the phase diagram and may even enable skyrmion stability far below the ordering temperature. Among cubic B20 helimagnets, the insulator Cu$_2$OSeO$_3$ is currently the only known example exhibiting a low-temperature, anisotropy-induced skyrmion pocket.
In this manuscript, we present a systematic study of cubic magnetocrystalline anisotropy by means of angle-resolved SQUID magnetization measurements in MnSi and Fe$_{1-x}$Co$_{x}$Si ($0.08 \leq x \leq 0.70$) single crystals and provide quantitative values of the anisotropy constants. For Fe$_{1-x}$Co$_{x}$Si, the cubic anisotropy is found to be strongly dependent on the Co concentration $x$. In particular, for low Co concentrations ($x\sim0.10$), the anisotropy is sufficiently strong to stabilize a low-temperature skyrmion lattice, in agreement with theoretical predictions. This finding suggests that Fe$_{1-x}$Co$_{x}$Si may represent the first chiral metallic system to exhibit a low-temperature skyrmion phase controllably stabilized by cubic anisotropy for specific directions of the magnetic field.
\end{abstract}

\maketitle

\section{\label{sec:level1}Introduction}

The $B$20 magnet MnSi has been established as a prototype chiral metal hosting a helimagnetic ground state and a rich magnetic phase diagram including the field-induced skyrmion lattice \cite{muehlbauer_2009, Neubauer09}. The principal interactions underlying the magnetic order in MnSi, \textit{i.e.}, a strong ferromagnetic exchange, intermediate antisymmetric exchanges, and weak cubic anisotropy, have been identified theoretically \cite{ishikawa1976, bak_jensen_1980, NAKANISHI_1980_MnSi}. Extensive experiments aimed at these principal interactions have resulted in a well-established generic phase diagram \cite{Grigoriev2006, Grigoriev06_MnSi_APhase, bauer_2012_MnSiSus, Bauer13}. MnSi undergoes a first-order phase transition into a helimagnetic ground state below $T_\text{HM} = 29.5~\text{K}$ \cite{Janoschek2013}. Applying an external field results in a transition from helical to conical ordering at the lower critical field $H_{c1}$. Further, above the higher critical field $H_{c2}$, the conical order completely field polarizes. Experiments show that for MnSi, the critical fields are only weakly dependent on the crystal orientation \cite{ishikawa1976, bauer_2012_MnSiSus}.

The isostructural doped semiconductor Fe$_{1-x}$Co$_{x}$Si has also been subject to many studies regarding the same principal interactions for specific Co concentrations \cite{Beille_1981, BEILLE_1983_fecosi, Motokawa1987, grigoriev2007, grigoriev2007_principal_interactions, muenzer_2010} and features the same generic phase diagram as MnSi \cite{grigoriev2007, grigoriev2007_principal_interactions, muenzer_2010, bauer2016_history, kindervater2019}. However, in contrast to MnSi, the distinctive feature of the Fe$_{1-x}$Co$_{x}$Si series is that key parameters such as ordering temperature, helical modulation length, spin-wave stiffness, Dzyaloshinskii–Moriya interaction strength, and cubic anisotropy can all be tuned systematically by varying the Co concentration, as demonstrated in a wide range of experiments \cite{Beille_1981, manyala_2000, Manyala2004, grigoriev2007, grigoriev2007_principal_interactions, Grigoriev2010, Grefe2024}. This tunability makes Fe$_{1-x}$Co$_{x}$Si an interesting platform for case studies, as opposed to MnSi, where the principal interactions are essentially fixed and can only be influenced by external parameters such as pressure \cite{BLOCH1975, Thessieu1998, Koyama2000, Ritz2013_MnSi_pressure}.

\begin{figure}
    \includegraphics[width=1.00\linewidth]{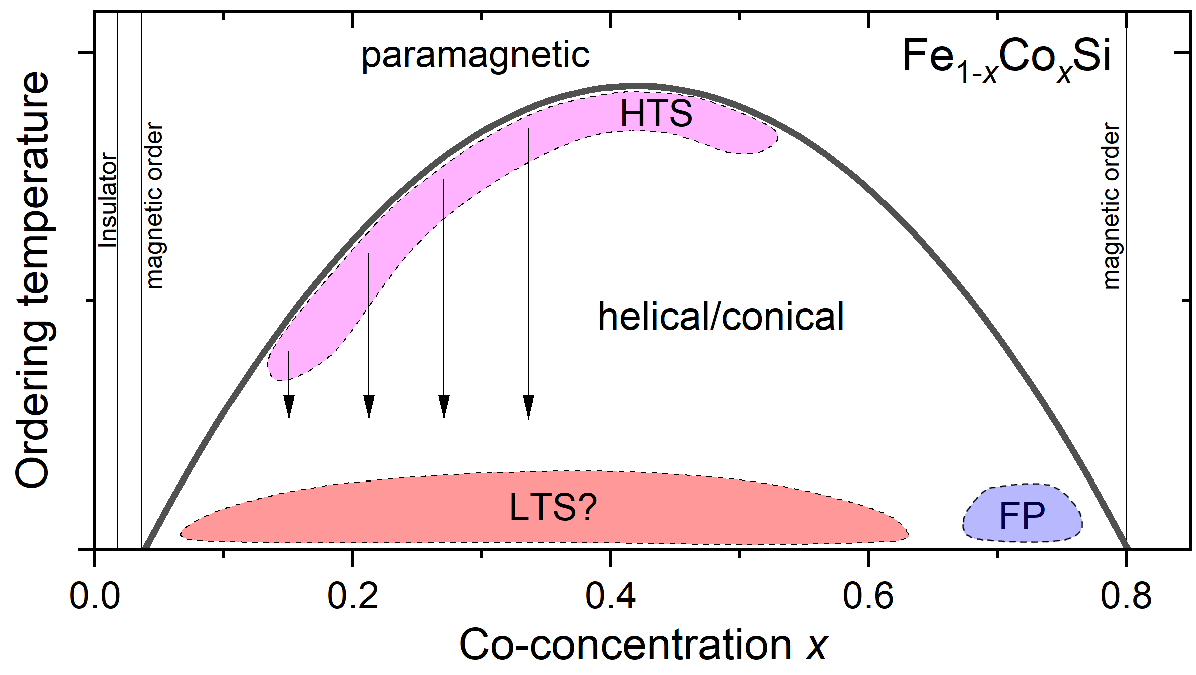}
    \caption{Schematic depiction of the phase diagram of Fe$_{1-x}$Co$_{x}$Si as function of the Co concentration $x$. At low Co concentrations, Fe$_{1-x}$Co$_{x}$Si undergoes an insulator-to-metal transition, closely followed by a nonmagnetic-to-magnetic transition \cite{Grefe2024}. For Co concentrations up to $x=0.50$, a high-temperature skyrmion phase (HTS) persists near $T_\text{HM}$ \cite{grigoriev2007, grigoriev2007_principal_interactions, muenzer_2010, bauer2016_history} with a strong history dependence \cite{bauer2016_history} as indicated by arrows. At high Co concentrations near $x\approx 0.65$ Fe$_{1-x}$Co$_{x}$Si shows a tendency to show full field polarization (FP) \cite{Grigoriev2022_FM}. From studies on Cu$_2$OSeO$_3$ \cite{2018Chacon, 2018Halder}, the question arises whether the tunable system Fe$_{1-x}$Co$_{x}$Si also features a low-temperature skyrmion phase (LTS).}
    \label{fig:schematic_phases}
\end{figure}

A schematic depiction of the phase diagram of Fe$_{1-x}$Co$_{x}$Si as a function of the Co concentration $x$ is shown in Fig.~\ref{fig:schematic_phases}. For $x=0.014$, the system undergoes a metal-to-insulator transition \cite{Grefe2024} from a correlated narrow-gap insulator featuring magnetic and conducting surface states \cite{Fang2018, Avers2024} to bulk metallic behaviour, closely followed by a nonmagnetic-to-magnetic phase transition at $x=0.03$ \cite{Grefe2024}. Up to $x\sim 0.35$ the ordering temperature $T_\text{HM}$ increases and exhibits a maximum at ${T_\text{HM}\approx 53~\text{K}}$. For high concentrations $x\sim 0.65$, a tendency to field polarize at very low fields has been reported \cite{Grigoriev2022_FM}, which is being attributed to vanishing asymmetric exchanges and finite cubic anisotropy. Above $x=0.80$, long-range magnetic order vanishes.

Another key feature of MnSi and Fe$_{1-x}$Co$_{x}$Si is the formation of a skyrmion lattice in a small region of the phase diagram near the ordering temperature \cite{muenzer_2010, bauer2016_history}, see Fig.~\ref{fig:schematic_phases}. Though there have been efforts to characterize the cubic anisotropy and its influence on the formation of the skyrmion lattice in MnSi \cite{2017Bannenberg, Lou2018, Adams2018} and related compounds \cite{kindervater2020, preissinger2021}, there exists no systematic experimental study regarding the anisotropy in MnSi and Fe$_{1-x}$Co$_{x}$Si. This is mostly due to the fact that the anisotropy in MnSi and Fe$_{1-x}$Co$_{x}$Si is weak and lowest in the energy hierarchy of the competing interactions, and has long been thought to only weakly orient the helix along the easy axes in zero field. Additionally, cubic anisotropy classically exhibits a strong temperature dependence such that the anisotropy vanishes much more rapidly than the ordered moment \cite{Akulov1936, CALLEN1966}. This makes it only more challenging to measure the anisotropy at temperatures in the vicinity of the fluctuation-induced skyrmion pocket in the phase diagram, which is referred to as the high-temperature skyrmion phase (HTS).

Recently, in the chiral insulating magnet Cu$_2$OSeO$_3$, a second skyrmion phase (LTS) has been identified at low temperatures far below the ordering temperature \cite{2018Chacon, 2018Halder, Bannenberg2019, 2023Crisanti}. Cu$_2$OSeO$_3$ shares similar properties with MnSi and Fe$_{1-x}$Co$_{x}$Si, such as the lack of inversion symmetry and the resulting antisymmetric magnetic exchange interactions, leading to the same principal phase diagram: a helimagnetic ground state, transitions into a conical and field-polarized state at $H_{c1}$ and $H_{c2}$, respectively, and a skyrmion phase near the ordering temperature \cite{2012Seki_science, 2012Seki_prb}. However, in contrast to the fluctuation-induced high-temperature skyrmion phase, the low-temperature skyrmion phase is stabilized by cubic anisotropy and exhibits a pronounced directional dependence \cite{2018Chacon, 2018Halder, Bannenberg2019, 2023Crisanti}, pointing out a crucial role of anisotropy at low temperatures. This observation raises the question of whether such a secondary skyrmion phase may also exist in metallic MnSi and Fe$_{1-x}$Co$_{x}$Si, highlighting the need for a detailed understanding of cubic anisotropy in these systems. For a systematic investigation, the series Fe$_{1-x}$Co$_{x}$Si provides a particular advantage, as parameters such as the cubic anisotropy can be tuned by varying the Co concentration. Further, a realization of an LTS phase in Fe$_{1-x}$Co$_{x}$Si would open the door for numerous new experiments aimed at the transport properties of anisotropy-stabilized skyrmions.

In Ref. \cite{2018Halder}, the anisotropy of Cu$_2$OSeO$_3$ is estimated by magnetization measurements. It is shown that the anisotropy affects the non-collinear magnetic order with opposing signs under certain conical angles, \textit{i.e.}, the magnetization curves measured along the easy and hard axes exhibit a cross-over \cite{2018Halder}. This leads to a situation in which the anisotropy of Cu$_2$OSeO$_3$ is not easily measured by extracting the magnetization energy from zero magnetization to saturation. For MnSi, the magnetization curves feature no crossover, and for Fe$_{1-x}$Co$_{x}$Si, these effects play only a negligible role as can be seen by the angular dependence of the critical fields $H_{c1}$ and $H_{c2}$, see below. Thus, for MnSi and Fe$_{1-x}$Co$_{x}$Si, the magnetic anisotropy may be determined by calculating the magnetization energy from angle-resolved magnetization measurements.

In this work, we not only aim at further narrowing down the field of uncharacterized magnetic interactions by measuring the cubic anisotropy of MnSi and Fe$_{1-x}$Co$_{x}$Si but also identify promising Co concentrations that could host an LTS phase. We conduct angle-resolved magnetization measurements to obtain the magnetization energies and determine the anisotropy constants by fitting the data with a phenomenologically motivated function considering the overall cubic symmetry of the system. We then discuss the method of measuring magnetic anisotropy using magnetization measurements for the system Fe$_{1-x}$Co$_{x}$Si by comparing the orientation dependence of the critical fields $H_{c1}$ and $H_{c2}$. Determining the anisotropy constants leads to an extended phase diagram, which shows that magnetic order and cubic anisotropy in Fe$_{1-x}$Co$_{x}$Si are not trivially connected. Putting anisotropy in the context of the critical field $H_{c2}$ and the saturation magnetization $M_s$ by calculating the unitless anisotropy constants, we show that relative anisotropy is strongest for Co concentrations where magnetic order emerges or vanishes, \textit{i.e.}, at the phase boundaries of the phase diagram. This method is complemented by the calculation of cubic anisotropy from theoretical simulations and values of $H_{c2}$. Comparison with a critical anisotropy threshold obtained from theoretical simulations leads to the identification of two promising Co concentrations by both aforementioned methods, likely to host a low-temperature skyrmion phase.

\section{\label{sec:level2}Experimental methods}

For the experiments single crystals of MnSi and Fe$_{1-x}$Co$_{x}$Si with Co concentrations $x$ = 0.08, 0.15, 0.20, 0.30, 0.50, 0.60, 0.65, and 0.70 are grown by the tri-arc Czochralski method from a premelt of 99.99\% Mn, 99.98\% Fe, 99.95\% Co, and Si ($\rho_n=300~\Omega\text{cm}$, $\rho_p=3000~\Omega\text{cm}$) in their respective stochiometric ratios. After growth, the crystals are Laue-oriented, and cube-shaped samples (${\sim1~\text{mm}^3}$) are cut such that the cube edges align with the $\langle 100 \rangle$ cubic main axes of the crystal.

\begin{figure}
    \begin{tikzpicture}
        \node[anchor=north west,inner sep=0] at (0,0) {\includegraphics[width=0.95\linewidth]{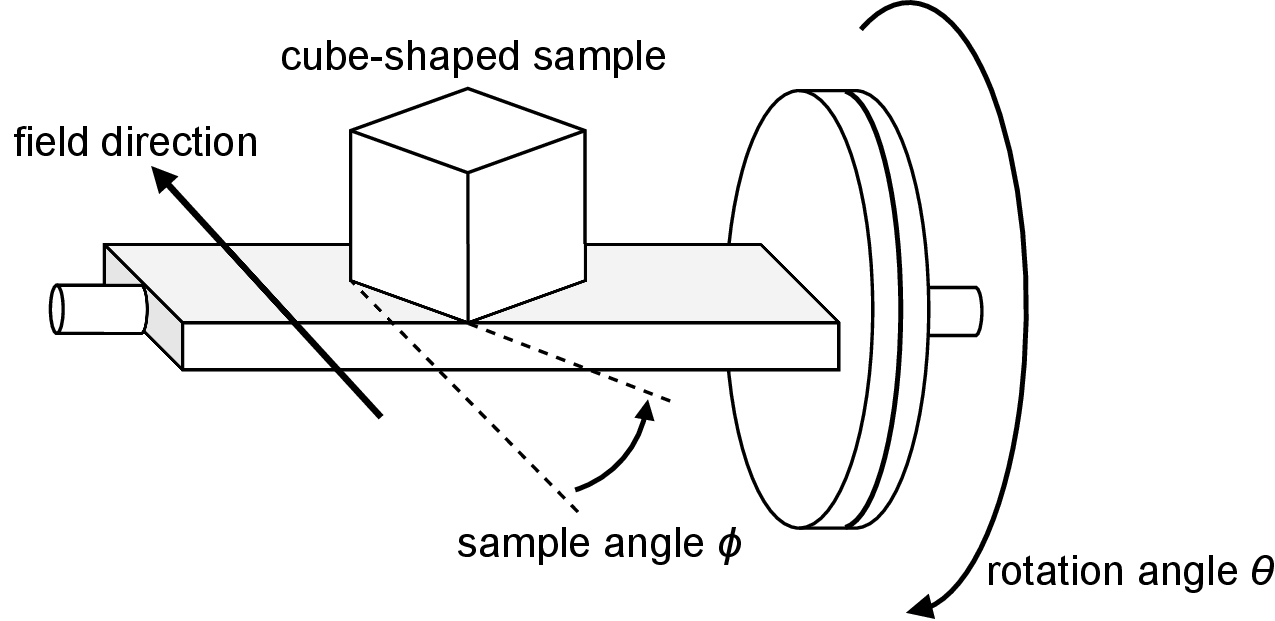}};
        \draw (0,0) node {\normalsize{(a)}};
        \node[anchor=north west,inner sep=0] at (0,-4.75) {\includegraphics[width=0.99\linewidth]{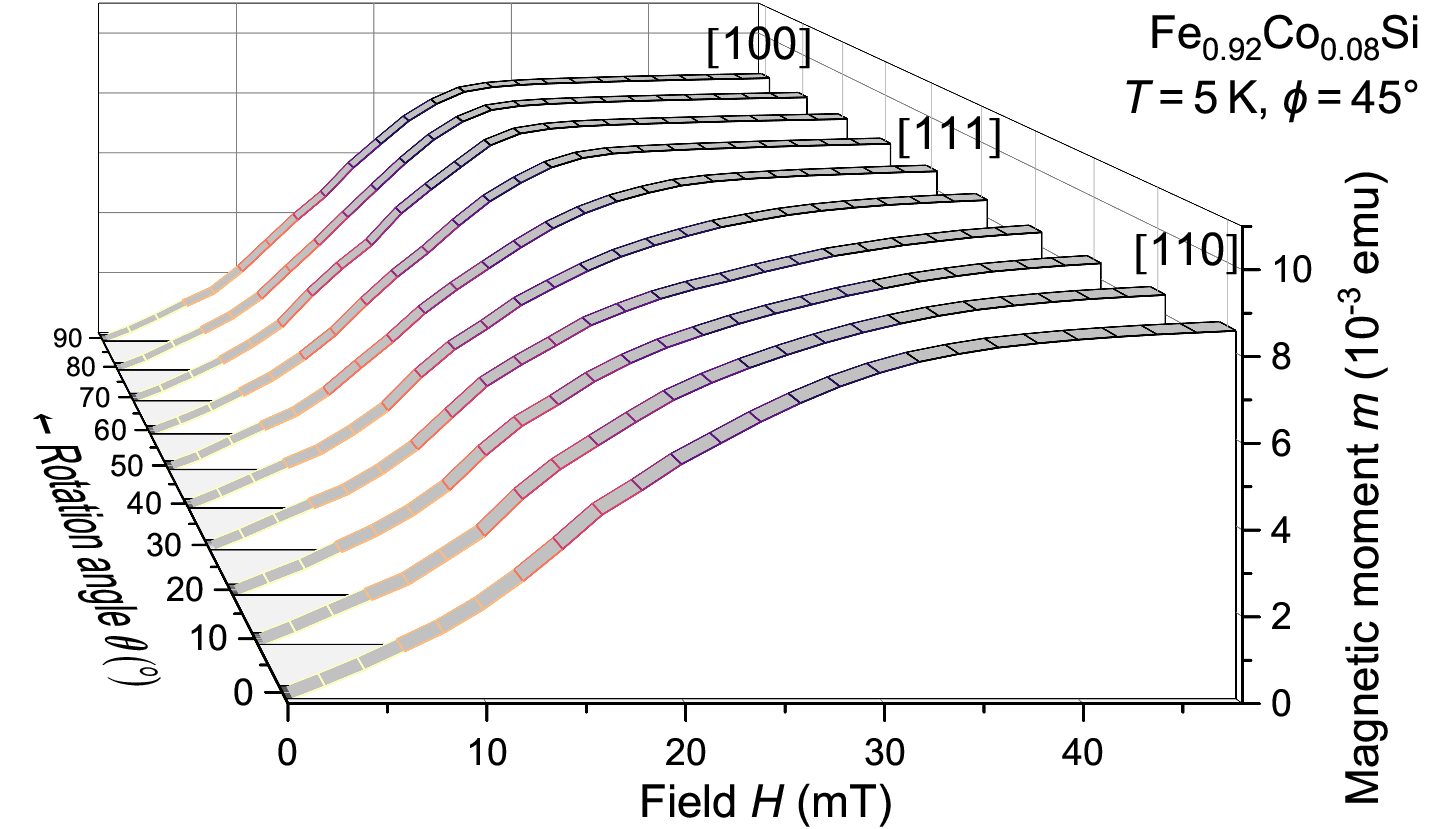}};
        \draw (0,-4.75) node {\normalsize{(b)}};
    \end{tikzpicture}
    
    \caption{(a) Rotatable sample holder used to conduct angle-resolved magnetization measurements and (b) exemplary measurements of the magnetic moment upon sample rotation for Fe$_{0.92}$Co$_{0.08}$Si at $T=5~\text{K}$. The magnetization direction relative to the crystal orientation is parameterized by the angles $\theta$ and $\phi$.}
    \label{fig:sampleholder}
\end{figure}

The magnetic anisotropy is measured by means of angle-resolved magnetization measurements using a rotatable sample holder in a commercial SQUID magnetometer. The direction of the crystal lattice relative to the magnetic field is described by the sample angle $\phi$ and the rotation angle $\theta$, see Fig.~\ref{fig:sampleholder} (a). The magnetization direction can be parameterized by a unit vector $\hat{\pmb{m}} = {\pmb M}/{|\pmb M|} = (m_1, m_2, m_3)$ in spherical coordinates where $\phi$ and $\theta$ are the azimuthal and the polar angle, respectively. The magnetization of all samples is measured up to saturation for multiple fixed sample angles and full rotations of the rotation angle in steps of $10^\circ$. Magnetization measurements are conducted for each rotation at $T=5~\text{K}$ and $T=10~\text{K}$ after heating beyond the respective ordering temperatures and cooling in zero-field to prevent history-dependent effects. Exemplary measurements for Fe$_{0.92}$Co$_{0.08}$Si at $T=5~\text{K}$ for $\phi = 45^\circ$ are shown in Fig.~\ref{fig:sampleholder} (b). 

To distinguish between intrinsic and extrinsic contributions to the angular dependence of the magnetization, a geometric offset in the magnetization caused by the sample holder and the coil setup is taken into account. This offset exhibits a twofold symmetry which can be distinguished from cubic anisotropy by symmetry arguments and is typically of the order between $10^{-3}M_s$ and $10^{-4}M_s$. Additionally, the demagnetization tensors $N$ of the different samples are calculated according to Ref. \cite{1998_aharoni} from the sample dimensions. The internal magnetic field is then given by $\pmb H = \pmb H_\text{ext} - N \cdot \pmb M$. 
The anisotropy can be characterized by calculating the magnetization energy $E$ per unit volume
\begin{eqnarray}
    E(\hat{\pmb m}) = \mu_0 \int_0^{M_\text{s}} \pmb{H} \cdot \text{d}\pmb{M}\,,
    \label{energy}
\end{eqnarray}
as a function of the magnetization direction $\hat{\pmb m}$ from the magnetization curve. As the $B$20 structure is cubic, the symmetry of the intrinsic magnetocrystalline anisotropy is also expected to be cubic. Here, we acknowledge that the local symmetry of the $B$20 unit cell is actually governed by tetrahedral symmetry; however, since the magnetization depends not strictly on a certain direction but rather on a respective directional axis, we consider overall cubic symmetry.

In a classical energy expansion for homogenously magnetized systems, cubic anisotropy is phenomenologically expressed in powers of direction components $m_1, m_2, m_3$ \cite{VanVleck1937}, such that the anisotropy energy can be described by 
\begin{eqnarray}\label{eq:cubic_anisotropy}
    E_\text{cub}(\hat{\pmb m}) = K_0 + K_1 \left(m_1^2m_2^2+m_2^2m_3^2+m_1^2m_3^2\right) + \text{...}\,.\;\quad
\end{eqnarray}

As the cubic anisotropy of the B20 magnets is generally weak, the first non-vanishing term in the expansion is sufficient to describe the data, see below. Expansion terms of higher orders are neglected. The anisotropy is then solely characterized by the sign and magnitude of the fourth-order anisotropy constant $K_1$. The second-order anisotropy constant $K_0$, which is constant for cubic symmetry, includes all other isotropic contributions to the magnetization curve.

\section{\label{sec:level3}Experimental results}

\begin{figure}
    \includegraphics[width=1\linewidth]{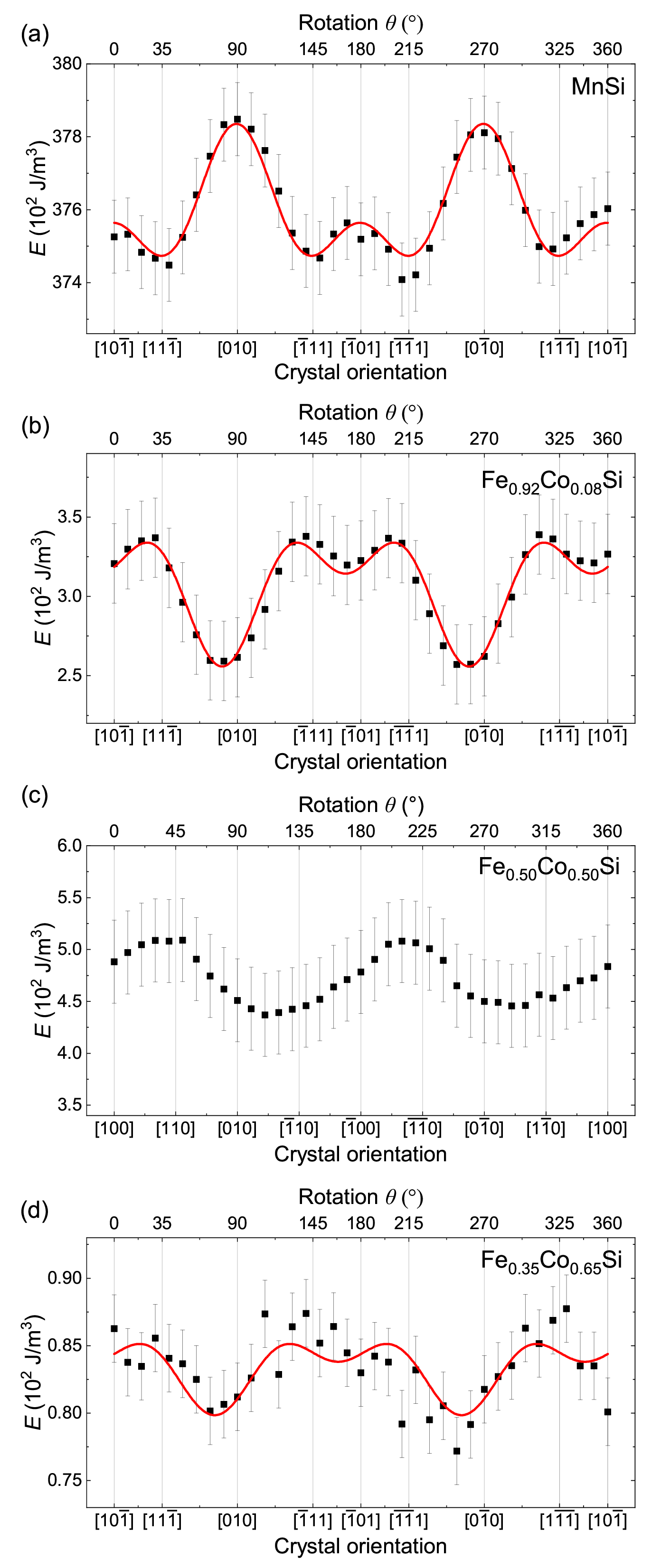}
    \caption{Magnetization energies $E$ as a function of the crystal orientation at $T=5~\text{K}$ for (a) MnSi with $\phi=45^\circ$, (b) Fe$_{0.92}$Co$_{0.08}$Si with $\phi=45^\circ$, (c) Fe$_{0.50}$Co$_{0.50}$Si with $\phi=0^\circ$ and (d) Fe$_{0.35}$Co$_{0.65}$Si with $\phi=45^\circ$. Red lines represent the cubic anisotropy fit according to Eq. (\ref{eq:cubic_anisotropy}). (a) represents the case of $K_1<0$, (b) and (d) the case of $K_1>0$, and (c) $K_1=0$ within the resolution of our experiment.}
    \label{fig:ergs_111}
\end{figure}

The magnetization energies $E$ calculated from the magnetization curves according to Eq. (\ref{energy}) as a function of the crystal orientation for MnSi, Fe$_{0.92}$Co$_{0.08}$Si, Fe$_{0.50}$Co$_{0.50}$Si, and Fe$_{0.35}$Co$_{0.65}$Si at selected sample angles are shown in Fig.~\ref{fig:ergs_111}.
A general feature of all samples under investigation, except Fe$_{0.50}$Co$_{0.50}$Si, which will be discussed separately, is that all principal axes are energetically equivalent within the resolution of our experiment. This generally confirms cubic symmetry, \textit{i.e.}, the magnetization energies are the same for each $\langle 100 \rangle$ direction and further motivates the procedure of data fitting with Eq.~(\ref{eq:cubic_anisotropy}) in cubic symmetry introduced in Sec. \ref{sec:level2}.

Moreover, the evolution of the orientation-dependent magnetic behaviour of MnSi and Fe$_{1-x}$Co$_{x}$Si can be divided into four cases:

(i) For MnSi, the anisotropy constant $K_1$ is negative. The easy axes and, thus, minima in magnetization energy are along the $\langle 111\rangle$ directions (Fig.~\ref{fig:ergs_111} (a)). This is in agreement with literature as the easy axes have been inferred from neutron scattering data \cite{Grigoriev2006}. In zero field, the spin-helix vectors are aligned with the $\langle 111\rangle$ directions.

(ii) For the low-doping regime in Fe$_{1-x}$Co$_{x}$Si ($x\leq0.30$) the anisotropy constant $K_1$ is positive. The easy axes are the $\langle 100 \rangle$ directions (Fig.~\ref{fig:ergs_111} (b)). This is consistent with the helix vector in zero field aligned along the $\langle 100\rangle$ directions as derived from neutron scattering experiments \cite{grigoriev2007}.

(iii) With increasing Co concentration, the anisotropy vanishes for Fe$_{0.50}$Co$_{0.50}$Si within the experimental resolution. Fig. \ref{fig:ergs_111} (c) shows the magnetization energy upon rotating covering two $\langle 100\rangle$ directions ($[100]$ and $[010]$) and two $\langle 110\rangle$ directions ($[110]$ and $[\overline 110]$).  For finite cubic anisotropy, each $\langle 100\rangle$ direction and each $\langle 110\rangle$ direction are expected to be energetically equivalent, exhibiting a four-fold symmetry. This is observed in our experiment for MnSi and other concentrations for Fe$_{1-x}$Co$_{x}$Si. In contrast, this is not the case for Fe$_{0.50}$Co$_{0.50}$Si, indicating vanishing cubic anisotropy, \textit{i.e.}, $K_1$ being zero. The residual orientation dependence is attributed to uncompensated demagnetization fields, which, however, can be distinguished from cubic anisotropy by the twofold symmetry.

(iv) In the high-doping regime for Fe$_{1-x}$Co$_{x}$Si (${x\geq0.60}$), the cubic anisotropy recovers to finite values of $K_1$ (Fig.~\ref{fig:ergs_111} (d)). However, the magnetization energies exhibit a far less pronounced orientation dependence than in the low-doping regime. The angular dependent magnetization energies are one order of magnitude weaker than for the low-doping regime. The data for all other Co concentrations with non-vanishing anisotropy is reminiscent of that of Fe$_{0.92}$Co$_{0.08}$Si, where only the magnitudes and, thus, the strength of the anisotropy, vary. In the high-doping regime, the energy scales of cubic anisotropy and demagnetization corrections become comparable, which leads to a reduced signal-to-noise ratio compared to the low-doping regime.

For a quantitative characterization of the magnetic anisotropy, the values of the anisotropy constants are determined by fitting the data according to Eq.~(\ref{eq:cubic_anisotropy}) up to fourth order. To compensate for a slight misorientation of the rotation angle $\theta$ when initializing a measurement, the data is fitted considering a rotational offset $\theta_0$ in the order of few degrees. The parameters $K_0$ and $\theta_0$ are set as free fit parameters for each orientation $\phi$. The parameter $K_1$ is a shared fit parameter across different orientations $\phi$ for each sample and temperature. 

Generally, there is no reason for the anisotropy to be solely cubic, to be limited to only fourth order or feature terms, including a helix vector-orientation dependence. However, since all our measurements for different orientations at different temperatures show the equivalence of all $\langle 111 \rangle$, $\langle 110 \rangle$, and $\langle 100\rangle$ directions, respectively, the inclusion of other expansion terms or fitting the data in sixth-order is not motivated by our data. Additionally, fitting with more parameters only marginally improves the fit at the expense of an additional free parameter. Hence, we choose to neglect terms not included in Eq. \eqref{eq:cubic_anisotropy}. 

Moreover, we emphasize that this approach does not resolve the individual contributions from spiral reorientation processes near $H_{c1}$ and $H_{c2}$, possible skyrmion phases, or multidomain effects, all of which can influence the magnetization curves. Instead, the procedure effectively provides an average magnetic energy over these different spin configurations by using Eq. (\ref{eq:cubic_anisotropy}). Consequently, the extracted value of $K_1$ should be interpreted as a first-order estimate of the cubic anisotropy. A more accurate determination requires a comprehensive theoretical treatment that explicitly accounts for the actual spin textures, which will be discussed in subsequent analysis.

\begin{figure}
    \includegraphics[width=1\linewidth]{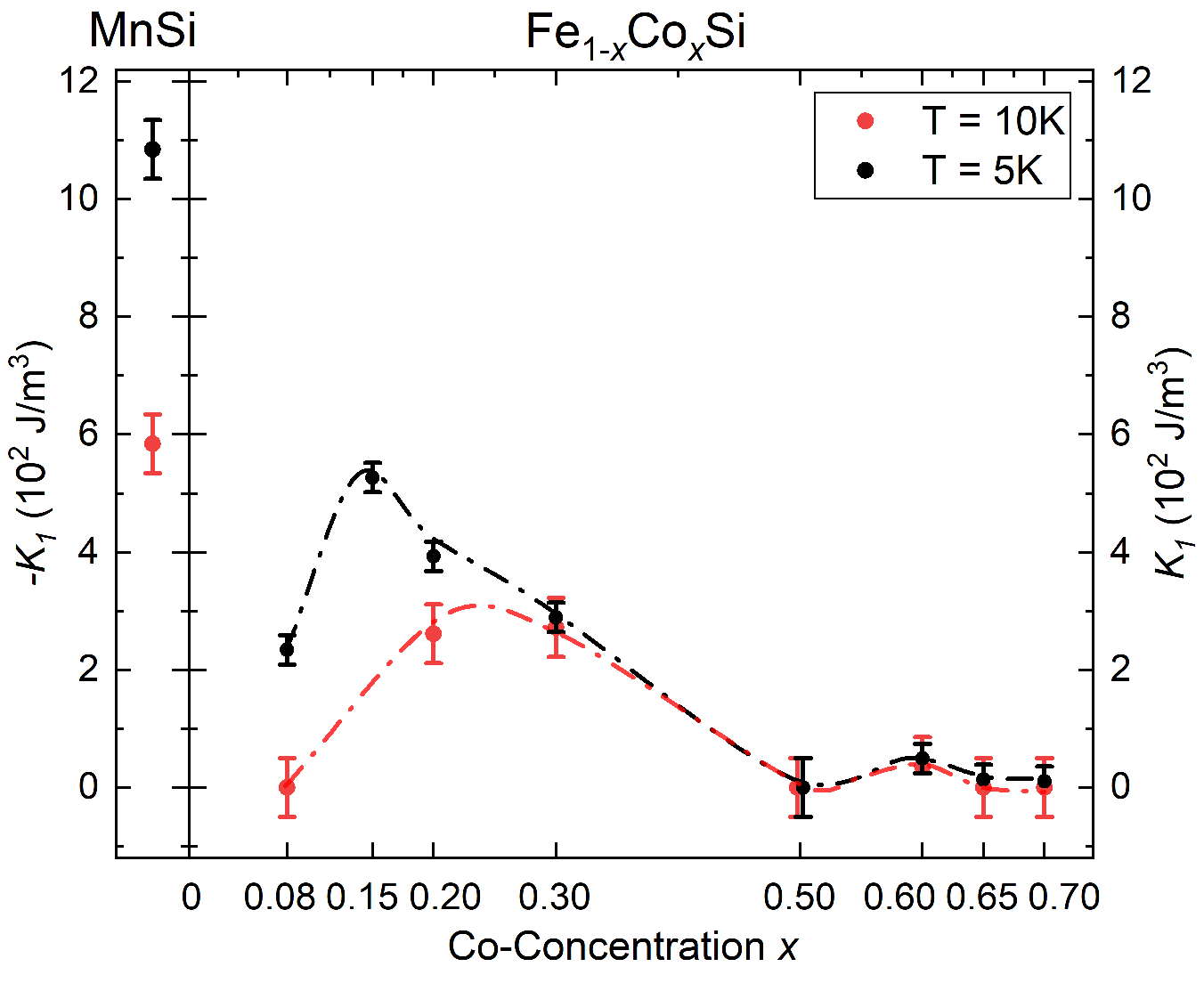}
    \caption{Anisotropy constants $-K_1$ for MnSi and $K_1$ for Fe$_{1-x}$Co$_{x}$Si ($0.08 \leq x \leq 0.70$) at $T=5~\text{K}$ and $T=10~\text{K}$. Dashed lines are guides to the eye.}
    \label{fig:K_aktuell}
\end{figure}

With the anisotropy constants $K_1$ derived at $T=5~\text{K}$ and $T=10~\text{K}$ for MnSi and Fe$_{1-x}$Co$_{x}$Si a new phase diagram is obtained (Fig.~\ref{fig:K_aktuell}). The total anisotropy is more pronounced for MnSi than for Fe$_{1-x}$Co$_{x}$Si at $T=5~\text{K}$ and $T=10~\text{K}$. For the system Fe$_{1-x}$Co$_{x}$Si, the general evolution shown in Fig. \ref{fig:ergs_111} is reflected in the anisotropy phase diagram, see Fig. \ref{fig:K_aktuell}. In the low-doping regime, the cubic anisotropy is finite with $K_1>0$. For Fe$_{0.50}$Co$_{0.50}$Si, however, the cubic anisotropy vanishes, while in the high doping regime, the cubic anisotropy is finite with $K_1>0$ but about one order of magnitude weaker than for low doping. The highest total anisotropy at $T=5~\text{K}$ is observed for Fe$_{0.85}$Co$_{0.15}$Si.

The total values of the absolute anisotropy in MnSi and Fe$_{1-x}$Co$_{x}$Si are more than one order of magnitude lower than in other skyrmionic compounds such as FeGe \cite{1970Ludgren_FeGe}, Cu$_2$OSeO$_3$ \cite{2017Stasinopoulos}, or (Co$_{0.5}$Zn$_{0.5}$)$_{20-x}$Mn$_x$ \cite{preissinger2021}. The values of $K_1$ for MnSi are consistent with values reported in Ref. \cite{Sauther21PfleidererDiss} when converting their results to our notation of Eq.~(\ref{eq:cubic_anisotropy}) as $K_1=-2K_1'$. Slight discrepancies are due to the use of a different experimental technique. In addition, data fitting with higher order expansion terms in Eq. \ref{eq:cubic_anisotropy} may lead to deviations as the expansion terms are non-orthogonal.

\section{\label{sec:level4}Discussion}
\subsection{Determination of cubic anisotropy from magnetization curves}
\begin{figure}
    \includegraphics[width=0.97\linewidth]{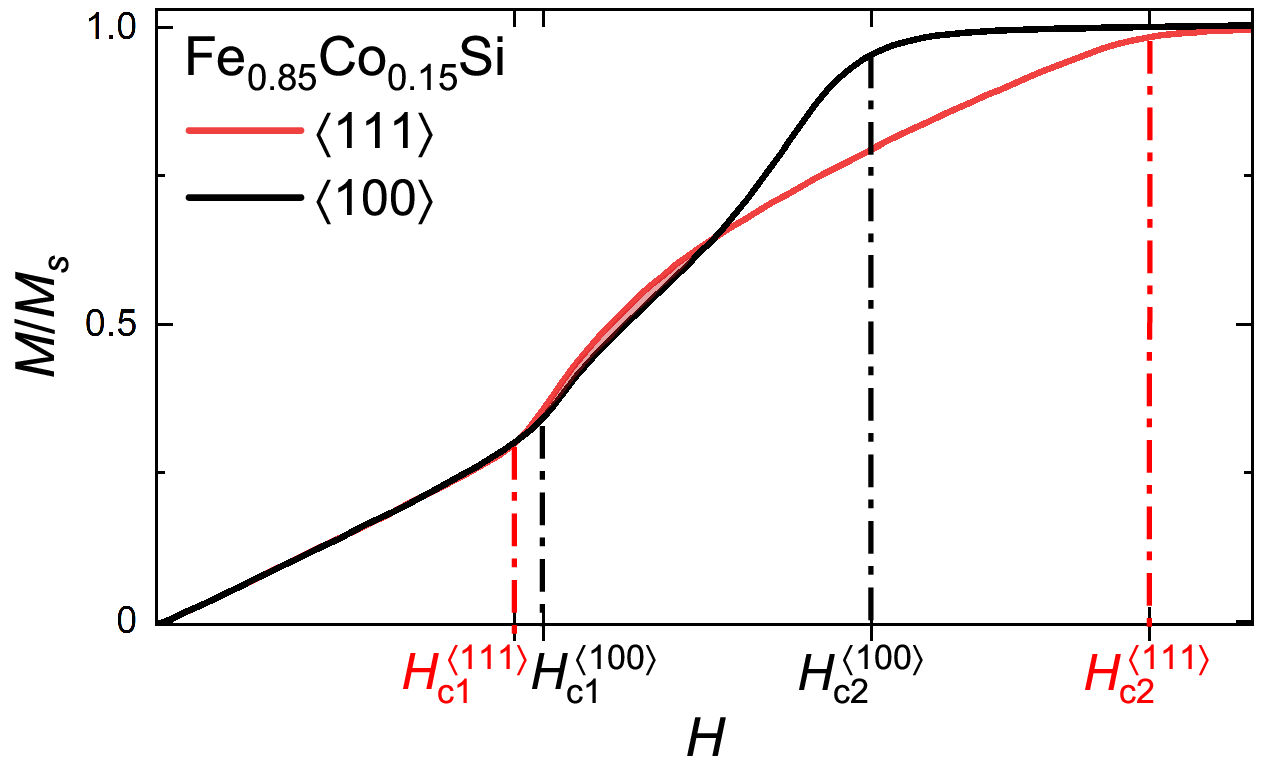}
    \caption{Magnetization $M$ for Fe$_{0.85}$Co$_{0.15}$Si at $T=5~\text{K}$ along the $\langle 100 \rangle$ and $\langle 111\rangle$ directions. The red-shaded region marks the area contributing with an opposite sign to the magnetic anisotropy. The critical fields are obtained from the first derivative and through linear interpolations near saturation.}
    \label{fig:overviewHC}
\end{figure}

Firstly, we want to discuss the validity of our method of determining the anisotropy constants from magnetization energies. Cubic anisotropies may be determined by a set of experiments, such as torque-magnetometry \cite{1970Ludgren_FeGe} or ferromagnetic resonance \cite{preissinger2021}, all of which have some sort of drawback. However, magnetization measurements present the most readily available technique. But, as mentioned above, using Eq. \eqref{eq:cubic_anisotropy} for an inhomogeneous magnet implies a form of averaging all spin structures, such as helicoids and skyrmions, from zero field up to saturation in a homogeneous approximation. This has the inherent flaw of not considering the formation of spin structures or reorientation processes, which also affect the magnetization curve. For example, as has been shown in Ref. \cite{2018Halder}, the critical fields in the system Cu$_2$OSeO$_3$ vary strongly with respect to the easy and hard axes. The magnetization along the easy and hard axes features a crossover between $H_{c1}$ and $H_{c2}$, \textit{i.e.}, the magnetization is steeper for low fields along the hard axis, but the saturation occurs at lower fields along the easy axis. Thus, measuring the magnetization energy does not yield reliable values for the anisotropy constants in Cu$_2$OSeO$_3$. For Fe$_{1-x}$Co$_{x}$Si, however, these effects are less pronounced. As shown in Fig. \ref{fig:overviewHC}, the magnetization of Fe$_{0.85}$Co$_{0.15}$Si does show a crossover between $H_{c1}$ and $H_{c2}$ for the easy and hard axes but the crossover region is much smaller than the difference induced by the cubic anisotropy. The crossing region contributes with an opposing sign to the net measured anisotropy. By calculating the contributing area in the magnetization curve, the error can be estimated to be around 1-2~\%. 

\begin{figure}
    \includegraphics[width=1\linewidth]{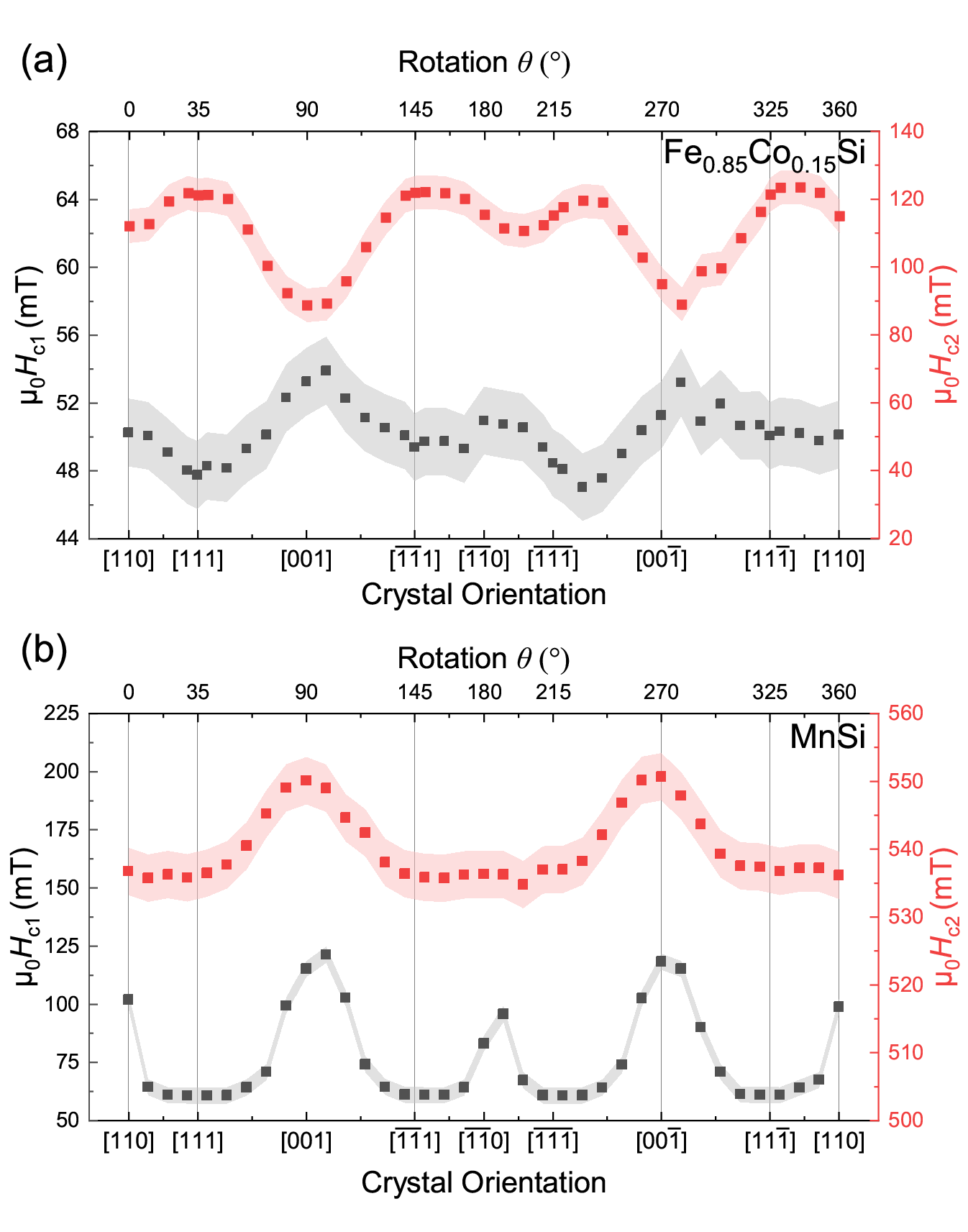}
    \caption{Critical fields $H_{c1}$ and $H_{c2}$ as a function of the crystal orientation at $T=5~\text{K}$ for (a) Fe$_{0.85}$Co$_{0.15}$Si and (b) MnSi. Values for $H_{c1}$ and $H_{c2}$ have been corrected by subtracting the projected demagnetization field. Shaded regions are errors governed by finite field increments during the measurements.}
    \label{fig:hc_111}
\end{figure}

This crossover can be visualized by the directional dependence of the critical fields for Fe$_{0.85}$Co$_{0.15}$Si, shown in Fig.~\ref{fig:hc_111} (a). For Fe$_{1-x}$Co$_{x}$Si at low Co concentrations ($x \leq 0.20$), the critical field $H_{c1}$ is maximal along the easy axes and shows a minimum along the hard axes. The critical field $H_{c2}$ shows the exact opposite behaviour, which leads to a crossover in magnetization shown in Fig~\ref{fig:overviewHC}. For higher concentrations $x \geq 0.30$, this crossover in magnetization is not observable in our measurements as the critical field $H_{c1}$ becomes essentially isotropic. For MnSi, however, there does not exist a crossover in the magnetization, and the critical fields show the same angular dependence as shown in Fig. \ref{fig:hc_111} (b). From these findings, it can be assumed that at least for Fe$_{1-x}$Co$_{x}$Si ($x \leq 0.20$), our values for $K_1$ are slightly underestimated, but the measurements for higher concentrations and MnSi are essentially unaffected by these effects. Thus, although being approximate, this approach yields a first-order estimate of the cubic anisotropy. The validity of the method is again reflected by our values of the anisotropy constants for MnSi being close to values reported in Ref. \cite{Sauther21PfleidererDiss} using a different experimental technique that is unaffected by these features in the magnetization.

\subsection{Co concentration dependence and low-temperature skyrmions in Fe$_{1-x}$Co$_{x}$Si}

\begin{figure}
    \includegraphics[width=0.99\linewidth]{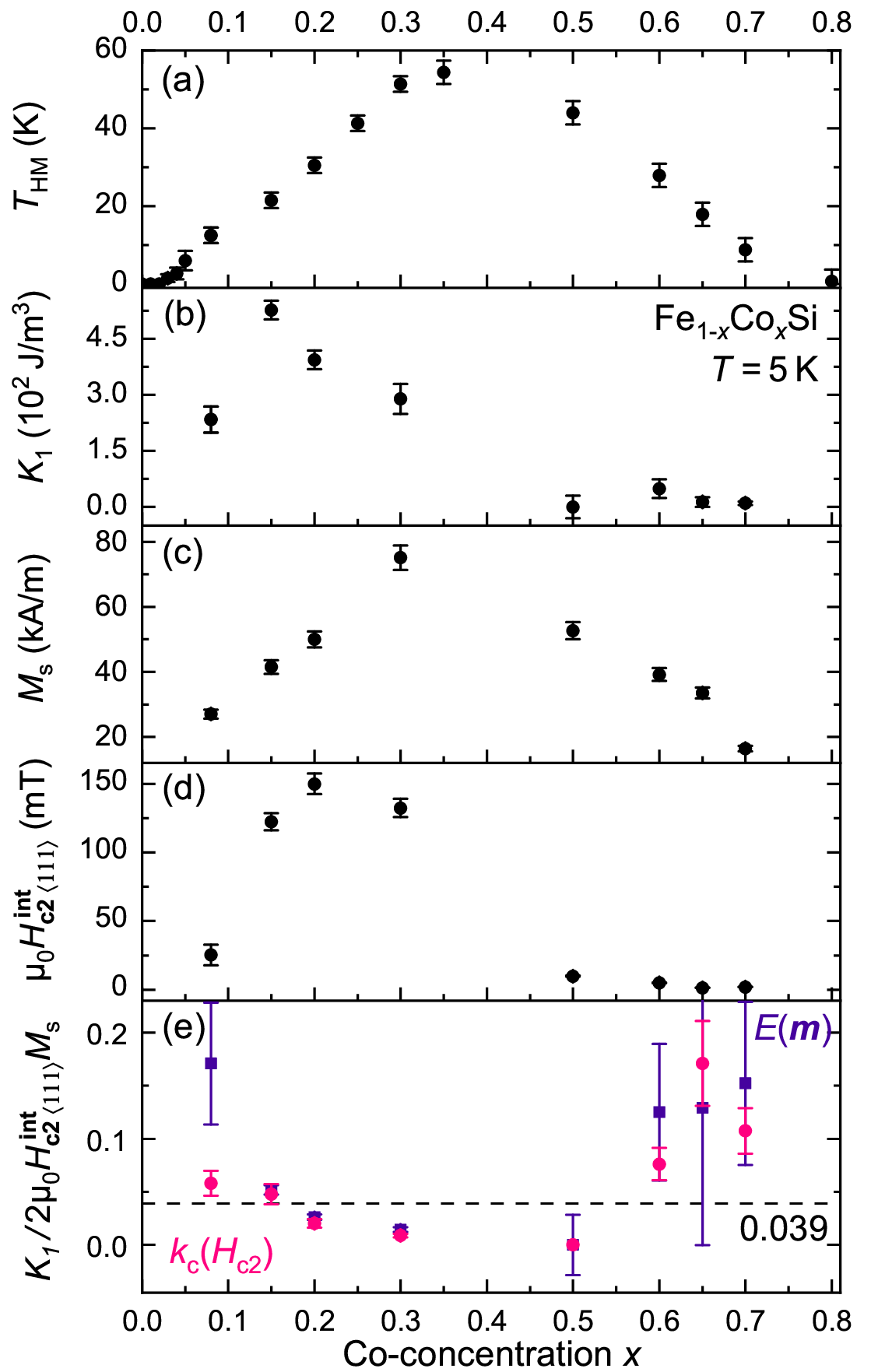}
    \caption{(a) Magnetic ordering temperatures $T_\text{HM}$, (b) anisotropy constant $K_1$, (c) saturation magnetization $M_s$, (d) internal critical field $H_{c2 \langle111\rangle}^\text{int}$, and (e) unitless anisotropy $k_c$ from magnetization curves (purple symbols) and $H_{c2}$ (pink symbols) at $T=5~\text{K}$ for Fe$_{1-x}$Co$_{x}$Si as a function of the Co concentration $x$. The threshold for LTS stabilization at $k_c=0.039$ according to Ref. \cite{Leonov2025_theory} is highlighted.}
    \label{fig:K_crit}
\end{figure}

To put our findings into the context of magnetic order in the system Fe$_{1-x}$Co$_{x}$Si, the ordering temperatures and anisotropy constants at $T = 5~\text{K}$ across the whole magnetic doping range are summarized in Fig. \ref{fig:K_crit} (a) and (b). The ordering temperature and anisotropy constant both feature maxima that do not coincide, and the evolution of the anisotropy as a function of the concentration does not resemble the trend of the ordering temperatures. We conclude that magnetic order and cubic anisotropy are not trivially dependent on each other. This can be easily understood when considering the respective energy scales. The values of the ordering temperatures are of the order of tens of Kelvin, resulting in an order of a few meV of magnetic exchange energies. Compared to that, if we consider the total cubic anisotropy per metal atom, we obtain an approximate energy of the order $10^{-4}$~meV, again emphasizing the stark difference in energy scales. Further, for Fe$_{1-x}$Co$_{x}$Si, the non-monotonic evolution of the anisotropy as a function of $x$ is best explained by the itinerancy of the metallic system. In itinerant magnets, anisotropy is strongly dependent on band-filling and, thus, doping \cite{Skomski2010}.

Though anisotropy and magnetic order seem to be essentially decoupled, it is sensible to consider the strength of the anisotropy relative to physical quantities defining the magnetic order. As motivated in Refs. \cite{2018Chacon, 2018Halder, Leonov2025_theory} the unitless anisotropy constant can be a measure for which thresholds of low-temperature skyrmion stability can be calculated theoretically. Here, we want to focus on the comparison between theoretical simulations giving a threshold of a critical dimensionless anisotropy constant of $k_c = K_cA/D^2= 0.039$ \cite{Leonov2025_theory} based on the phenomenological Dzyaloshinskii model, see for example \cite{2020Leonov, Roessler_2011}. From the framework given in Ref. \cite{Leonov2025_theory}, the dimensionless anisotropy constant $k_c$ can be related to the critical field and the saturation magnetization as 
\begin{equation}\label{eq:Kc}
    k_c = K_c\frac{A}{D^2} = \frac{K_c}{\mu_0 H_DM} = \frac{K_c}{2\mu_0 H_{c2}M}~,
\end{equation}
such that to compare the experimental findings to the theoretical prediction for $k_c$, we calculate $K_1/2\mu_0H_{c2 \langle111\rangle}^\text{int}M_s$ from our experiments. For the critical field, we again consider the internal magnetic field where the respective projected demagnetization field has been subtracted. The saturation magnetization, critical fields, and unitless anisotropy constants are shown in Fig.~\ref{fig:K_crit} (c), (d), and (e) (purple symbols), respectively. The threshold of $k_c = 0.039$ is indicated as a dashed line in Fig.\ref{fig:K_crit} (e). 

The values of $H_{c2}$ and $M_s$ are in good agreement with literature \cite{bauer2016_history, manyala_2000, Manyala2004, grigoriev2007, grigoriev2007_principal_interactions, Grigoriev2010}. And when considering the unitless anisotropy as a function of the Co concentration, it is apparent that well within the phase boundaries of magnetic order, the magnetic anisotropy is very weak compared to the other magnetic exchanges, represented by the value of the unitless anisotropy being close to zero. Interestingly, however, near the phase boundaries where magnetic order emerges or vanishes, the anisotropy gains considerably relative to $H_{c2}$ and $M_s$. At the boundaries, that is, very low or high doping, magnetic anisotropy at low temperatures becomes comparable to the other magnetic exchange interactions.  

In the low-doping regime, the samples with $x = 0.08$ and $0.15$ show significant dimensionless anisotropy, enough to stabilize skyrmions at low temperatures. Considering experimental requirements on temperature reachability and field accuracy, \textit{i.e.}, maximizing $H_{c2}$ and $T_\text{HM}$, Fe$_{0.85}$Co$_{0.15}$Si presents the most promising platform to search for the LTS phase at low temperatures.

Regarding the high-doping regime, it should be noted that the errors at high concentrations are governed by the aforementioned worsened signal-to-noise ratio in magnetization energies leading to errors in values of $K_1$, and because of the low values of $H_{c2}$ along the hard axes, demagnetization corrections are comparable to the values of $H_{c2}$. Because of the very low $H_{c2}$ fields there exist similarities to the situation reported for the related compound Fe$_{1-x}$Co$_{x}$Ge \cite{grigoriev2015}, where, at a certain critical concentration, the Dzyaloshinskii-Moriya interaction vanishes and the combination of ferromagnetic exchange with cubic anisotropy results in full field polarization. Compared to literature, the gain in relative anisotropy at high Co concentrations for Fe$_{1-x}$Co$_{x}$Si may be the cause of the tendency to field-polarize at very low fields, as inferred from neutron scattering data \cite{Grigoriev2022_FM}, indicating a vanishing antisymmetric contribution and the interplay of ferromagnetic exchange and strong cubic anisotropy. From the experimental point of view, finding low-temperature skyrmions in a field range comparable to the internal $H_{c2}$ of a few mT seems unlikely in comparison to the low Co concentrations.

\subsection{Anisotropy determined from numerical simulations and $H_{c2}$}
As has been mentioned above, the method of extracting quantitative values of the anisotropy by magnetization energies does not fully take contributions to the magnetization curve from the different non-homogeneous spin states into account. Especially when trying to predict low-temperature skyrmion stability in a narrow parameter range, these deviations might play a role such that the magnetization energy approach should not be the sole method for determining anisotropies. This motivates a more thorough approach from the theoretical side, where domain effects and reorientation processes can be considered. For that, again, the phenomenological Dzyaloshinskii model can be deployed to numerically simulate the critical field $H_{c2}$ depending on the unitless anisotropy constant $k_c$ and calculating the parameter $\Delta$. For a detailed description of the methods described, see Ref. \cite{Leonov2025_theory}. 

\begin{figure}
    \includegraphics[width=1\linewidth]{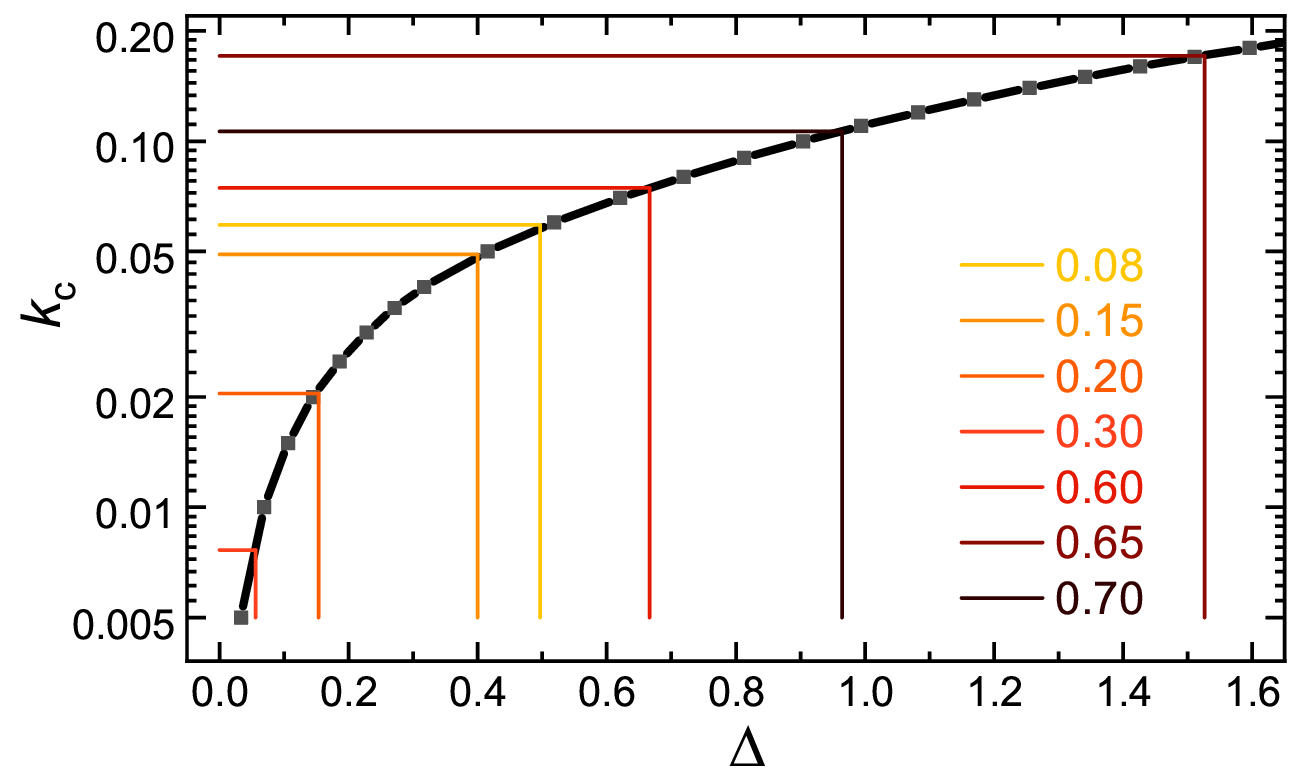}
    \caption{Dimensionless anisotropy $k_c$ for Fe$_{1-x}$Co$_{x}$Si as a function of the parameter $\Delta$ inferred from numerical simulations in the Dzyaloshinskii framework, see \cite{Leonov2025_theory}. Values for $\Delta$ and $k_c$ for the different Co concentrations are marked by solid lines. The sample with $x=0.50$ is not shown, as the critical field $H_{c2}$ is isotropic in the resolution of our experiments.}
    \label{fig:delta_kc}
\end{figure}
The parameter $\Delta$ offers the advantage that it can readily be calculated from experimental values of the critical fields for as
\begin{equation}
    \Delta = \dfrac{H_{c2[111]} - H_{c2[100]}}{H_{c2[100]}},
\end{equation}
for positive values of $k_c$, as is the case for the system Fe$_{1-x}$Co$_{x}$Si. From the simulations in Fig.~\ref{fig:delta_kc} and the calculated parameter $\Delta$ from $H_{c2}$, the unitless anisotropy constant can be determined, offering a method to extract $k_c$ from the angular dependence of $H_{c2}$ without the aforementioned uncertainty of the method utilizing magnetization energies. The values of the method utilizing $H_{c2}$ and numerical simulations are shown in Fig.~\ref{fig:K_crit} (e) as pink symbols.

Strikingly, the non-dimensional anisotropy parameters $k_c$ inferred from $H_{c2}$ and those from magnetization energies $E(\hat{\pmb m})$ coincide for most samples in our study. This highlights the applicability of both methods as they seem to yield similar results from two very different starting points, one being theoretical and the other experimental. Additionally, regarding the threshold of $k_c = 0.039$, both methods indicate the potential stability of low-temperature skyrmion phases for Fe$_{1-x}$Co$_{x}$Si at very low and very high Co concentrations. However, as discussed above, the findings for the high-doping regime are best interpreted as in Ref. \cite{Grigoriev2022_FM} and the experimental realisation of an LTS phase at such low critical fields seems unlikely.

However, there exist discrepancies for the $x=0.08$ sample for both methods, where the value determined by $E(\hat{\pmb m})$ exceeds the one based on $H_{c2}$. This most likely has two main contributions: for one, the values of $H_{c2}$ are small and only determined with some margin of error as demagnetization corrections become comparable. On the other hand, the saturation magnetization for Fe$_{1-x}$Co$_{x}$Si is not easily measured. For all samples, the magnetic moment does not fully saturate, even at moderately high fields \cite{Grefe2024, Onose_2005}. This becomes worse in the vicinty of the quantum critical point $x\approx0.03$ \cite{Grefe2024}, as is the case for the $x=0.08$ sample, leading to an underestimation of the saturation magnetization and thus an overestimation of $k_c$ calculated from Eq. \eqref{eq:Kc}.
\vspace{6mm}

\section{\label{sec:level5}Conclusions}

Summarizing our findings, we have shown that angle-resolved magnetization measurements present a viable technique to quantitatively measure total cubic anisotropies in itinerant systems such as MnSi and Fe$_{1-x}$Co$_{x}$Si. Further, with the use of this technique, we have established an anisotropy phase diagram for Fe$_{1-x}$Co$_{x}$Si across the whole magnetic doping range, having identified characteristic doping regimes. Putting the anisotropy in the context of magnetic order, cubic anisotropy seems to be decoupled from the ordered moment due to the steep energy hierarchy. When considering a unitless anisotropy relative to the exchange energies represented by $H_{c2}$ and the ordered moment $M_s$, we have identified regions with strong relative anisotropy at the phase diagram boundaries where magnetic order emerges/vanishes. This is in overall good agreement with dimensionless anisotropies calculated from values of $H_{c2}$ and numerical simulations. Compared to a theoretical threshold of $x_c=0.039$, the relative anisotropy for $x= 0.08$ and $0.15$ exceeds a critical value needed to stabilize a secondary skyrmion phase at low temperatures, again indicated by both values determined from magnetization curves and $H_{c2}$.

\begin{acknowledgments}
We acknowledge fruitful discussions with A. Rosch and U. K. Rößler.

\end{acknowledgments}
\begin{comment}

\appendix

\section{Appendixes}
\begin{table}[h!]
\caption{\label{tab:table1}
Anisotropy constants for MnSi and Fe$_{1-x}$Co$_{x}$Si in units of $10^2~\text{J}/\text{m}^3$.}
\begin{ruledtabular}
\begin{tabular}{ccc}
 & $K_1$ at $5~\text{K}$ & $K_1$ at $10~\text{K}$\\
\hline
MnSi & -10.85 & -5.85\\
Fe$_{0.92}$Co$_{0.08}$Si & 2.34 & 0\\
Fe$_{0.85}$Co$_{0.15}$Si & 5.27 & -\\
Fe$_{0.80}$Co$_{0.20}$Si & 3.93 & 2.61\\
Fe$_{0.70}$Co$_{0.30}$Si & 2.89 & 2.72\\
Fe$_{0.50}$Co$_{0.50}$Si & 0 & 0\\
Fe$_{0.40}$Co$_{0.60}$Si & 0.45 & 0.35\\
Fe$_{0.35}$Co$_{0.65}$Si & 0.13 & 0\\
Fe$_{0.30}$Co$_{0.70}$Si & 0.10 & 0\\
\end{tabular}
\end{ruledtabular}
\vspace{10px}
\end{table}

\end{comment}
\bibliography{anisotropy}% Produces the bibliography via BibTeX.

\end{document}